\documentclass[10pt,a4paper]{article}
\usepackage[utf8]{inputenc}
\usepackage{amsmath}
\usepackage{amsfonts}
\usepackage{amssymb}
\usepackage{url}
\usepackage{braket}
\usepackage{subfigure}
\usepackage{graphicx}
\usepackage[left=2cm,right=2cm,top=2cm,bottom = 2cm]{geometry}
\usepackage{xcolor}
\usepackage{lineno,hyperref}
\usepackage{floatrow}
\usepackage{booktabs}
\usepackage{multirow}
\modulolinenumbers[5]

\begin{document}

\title{Phase Shift Analysis of Light Nucleon-Nucleus Elastic Scattering using Reference Potential Approach}%\tnoteref{mytitlenote}}

\author{Lalit Kumar, Shikha Awasthi, Anil Khachi, O.S.K.S Sastri\\Department of Physical and Astronomical Sciences, Central University of Himachal Pradesh\\ Dharamshala, Himachal Pradesh-176215, Bharat(India)}

% or include affiliations in footnotes:
%\author[mymainaddress,mysecondaryaddress]{Department of Physical and Astronomical Sciences, Central University of Himachal Pradesh, Dharamshala, Himachal Pradesh-176215, Bharat(India)}

%\author[mysecondaryaddress]{Global Customer Service\corref{O.S.K.S Sastri}}
%\cortext[mycorrespondingauthor]{O.S.K.S Sastri}
%\textit{osks@hpcu.ac.in}

\maketitle
\begin{abstract}
The neutron and proton scattering with either deuteron or stable alpha particle can be modeled as a two particle system. In this paper, using Morse function as reference potential, inverse potentials have been computationally constructed directly from scattering phase shifts (SPS) data for light nucleon-nucleus systems.  The phase equation for various $\ell$-channels has been numerically solved using 5th order Runge-Kutta (RK-5) method in each iteration within the optimisation procedure to obtain best model parameters of Morse function that minimise mean absolute percentage error (MAPE). The inverse potentials for S-wave ($\ell=0$) of neutron-deuteron (\textit{nd}) and proton-deuteron (\textit{pd}) systems have been obtained with MAPE of 1.95 and 2.05$\%$ respectively. Those corresponding to various P and D channels of n-alpha ($n\alpha$) and p-alpha ($p\alpha$) systems, have been determined to less than 1.53 and 2.17$\%$ respectively. The obtained(experimental) resonance energies for p$_{1/2}$ and p$_{3/2}$, from their partial cross-sections plots, respectively are 4.1(4$\pm$1) and 0.93(0.89) for n-alpha and 5.21(5$\pm$2) and 1.96(1.96) for p-alpha system. While total cross-section for $n\alpha$ has been found to be matching with values available in literature, that of $p\alpha$ is seen to be following the correct trend.  
\end{abstract}


%\linenumbers

\section{Introduction}
Scattering is a phenomenon that is essential in understanding many nuclear properties. Numerous theoretical models have been developed to emphasise the underlying physics behind these scattering results. The scattering of light nuclei with nucleons and among themselves is crucial for understanding the underlying interaction and providing structural information. Several groups have conducted substantial research on both nucleon-deuteron \cite{1,2,3,4} and nucleon-alpha systems \cite{5,6,7,8,9,10}, with results of elastic scattering phase shifts reported both experimentally and theoretically.
In the cluster model \cite{11,12,13,14,15} description of nuclei, the study of nucleon-$\alpha$ elastic scattering as a two-body problem at low energy is important.  
Alpha-particles have a stable and symmetric structure and hence its deformation due to a collision with a nucleon can usually be ignored in the first approximation. Further, since it has no spin, there are no coupled equations in the collision analysis, and polarisation phenomena are limited to the nucleon. It contains no excited states below at least 20 MeV, therefore there will be no inelastic scattering to confound the understanding of collision processes at lower energies.

To study elastic scattering among light nuclei, generator coordinate method (GCM) or the resonating group method (RGM) were usually employed. For elastic scattering phase shifts (SPS), incorporation of phenomenological two-body interactions in this model results in strong agreement with experimental evidence \cite{16,17}. To examine nucleon-$\alpha$ elastic scattering below 18 MeV, Satchler et al. \cite{18} employed an optical potential model and Dohet-Early and Baye \cite{19} used the formalism of unitary correlation operator method and found good agreement with experimental results \cite{20}. Using the phase function method (PFM), Laha et. al., \cite{21,22,23} suggested a simple phenomenological potential model based on Manning-Rosen potential for nucleon-$\alpha$ elastic scattering.

Buck et al., \cite{24}, have argued that local potential can be considered for nucleon-nucleus and nucleus-nucleus systems without resorting to the resonating group method (RGM) \cite{25}.
They also came to the conclusion that non-local potential is primarily of the same type as local potential. Using a local Gaussian potential as the model of interaction, they were able to produce scattering phase shifts (SPS) for the $\alpha-\alpha$ and $\alpha-{}^{3}He$ systems with reasonable success. The model interaction obtained on solving the time independent Schrodinger equation (TISE) for Gaussian potential using matrix method \cite{26} in conjunction with variational Monte-Carlo (VMC) \cite{27} for $\alpha-\alpha$ system in SPS \cite{28} that match well with experimental data. We adopted a unique technique \cite{29} to get SPS, in which PFM is directly used in optimisation procedure to get model parameters that minimise mean absolute percentage error (MAPE) between simulated and experimental data. The efficacy of this method has been utilised to compile data for alpha-alpha scattering \cite{30} from various sources and performing MAPE analysis that resulted with resonant peaks in close match with experimental data. Recently, we have presented the phase function analysis of the alpha-nucleon systems using a Gaussian potential \cite{31}. Utilising the Morse function as zeroth reference potential \cite{32} to obtain the model of interaction directly from the experimental SPS, inverse potential for nucleon-deuteron systems \cite{33,34} for energy range upto 10 MeV have been obtained. In this paper, we extend the study to all available energies for nucleon-deuteron (\textit{nd}) and proton-deuteron (\textit{pd}) and obtain inverse potentials for n-alpha ($n\alpha$) and p-alpha ($p\alpha$) systems for low energies upto 18 MeV.
 
\section{Methodology}
The Morse function chosen as reference potential, is given by
\begin{equation}
V(r) = V_0\left(e^{-2(r-r_m)/a_m}-2e^{-(r-r_m)/a_m}\right)
\label{eq1}
\end{equation} 
where the model parameters $V_0$, $r_m$ and $a_m$ reflect strength of interaction, equilibrium distance at which maximum attraction is felt and shape of the potential respectively.\\
For $p\alpha$ \& \textit{pd} systems, Coulomb interaction that needs to be included is chosen as
\begin{equation}
V_{Coul(r)} = \frac{z_{1} z_{2} e^2}{r}erf(\beta r)
\end{equation}
for $p\alpha$, $\beta = 0.609 $fm$^{-1}$ \cite{35}, \textit{pd} $\beta = 0.441 $fm$^{-1}$ \\
The Schr$\ddot{o}$dinger wave equation for a particle undergoing scattering having energy E and orbital angular momentum $\ell$ is given by
\begin{equation}
\frac{\hbar^2}{2\mu} \frac{d^2u_{\ell}(k,r)}{dr^2}+\big[k^2-\ell(\ell+1)/r^2\big]u_{\ell}(k,r)=V(r)u_{\ell}(k,r)
\label{SchrEq}
\end{equation}
where $k=\sqrt{E/(\hbar^2/2\mu)}$, and $(\ell(\ell+1)\hbar^2)/{2\mu r^2}$ is centrifugal potential.
%The centrifugal potential is given by
%\begin{equation}
%V_{cf} = \frac{l(l+1)\hbar^2}{2\mu r^2}
%\end{equation}
Where, $\mu_{n\alpha} = \frac{m_{n} * m_{\alpha}}{m_{n} + m_{\alpha}}$ is the reduced mass  for $n\alpha$ and $\mu_{p\alpha} = \frac{m_{p} * m_{\alpha}}{m_{p} + m_{\alpha}}$ is the reduced mass  for $p\alpha$ with $m_{n} = 939.5654 MeV/c^2$, $m_{p} = 938.2720MeV/c^2$ and $m_{\alpha}  = 3728.7705 MeV/c^2$. And $\mu_{nd}= \frac{m_{n}m_{d}}{m_{n}+m_{d}}$, $\mu_{pd}= \frac{m_{p}m_{d}}{m_{p}+m_{d}}$
with $m_d$ = 469.4590 MeV/$c^2$ respectively.
 
This second order differential equation has been transformed to the following first order non-homogeneous differential equation of Riccati type \cite{36,37}, known as phase equation: 
\begin{equation}
\delta_{\ell}'(k,r)=-\frac{V(r)}{k}\bigg[cos(\delta_\ell(k,r))\hat{j}_{\ell}(kr)-sin(\delta_\ell(k,r))\hat{\eta}_{\ell}(kr)\bigg]^2
\label{PFMeqn}
\end{equation}
with initial condition $\delta_{\ell}(k,0) = 0$.

\begin{figure}[h!]
\caption{Flowchart showing implementation procedure for obtaining the inverse potential}
\centering
\includegraphics[width=0.7\textwidth]{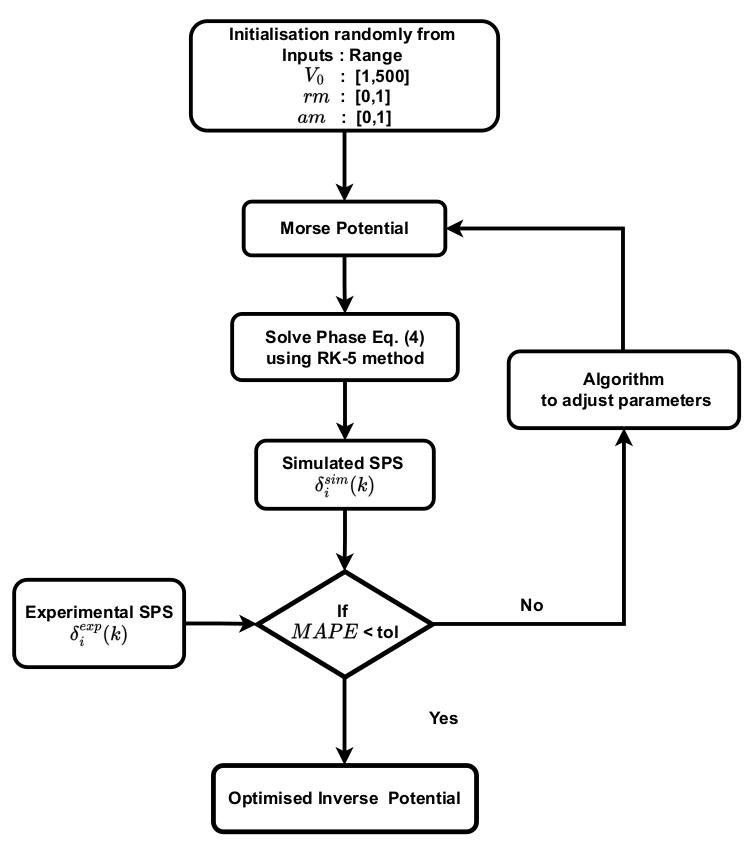}
\label{Flowchart} 
\end{figure}

The Riccati-Bessel \& Riccati-Neumann functions for different $\ell$ are obtained using
\begin{equation}
\hat{j}_{\ell}(kr)=(-kr)^{\ell}\bigg[\frac{1}{(kr)}\frac{d}{d(kr)}\bigg]^{\ell}\frac{sin(kr)}{(kr)}
\end{equation}
and
\begin{equation}
\hat {\eta_{l}}(kr)=-(-kr)^{\ell}\bigg[\frac{1}{(kr)}\frac{d}{d(kr)}\bigg]^{\ell} \frac{cos(kr)}{(kr)}
\end{equation}
respectively.
The phase equation for
\begin{enumerate}
\item[\textbf{(i)}] $\ell$ = 0, S-wave :
\begin{equation}
\delta'_0(k,r)=-\frac{V(r)}{k}sin^2[kr+\delta_0(k,r)]
\end{equation}
\item[\textbf{(ii)}] $\ell$ = 1, P-wave :
\begin{eqnarray}
\delta_1'(k,r)=-\frac{V(r)}{k}\bigg[cos(\delta_{\ell}(k,r))\bigg(\frac{sin (kr)}{kr}-cos(kr)\bigg)+\\ \nonumber sin(\delta_{\ell}(k,r))\bigg( \frac{cos(kr)}{kr}+sin(kr) \bigg)\bigg]^2
\end{eqnarray}
\item[and \textbf{(iii)}] $\ell$ = 2, D-wave :
\begin{equation}
\delta_2'(k,r) = -\frac{V(r)}{k}\big[cos(\delta_2(k,r))f(kr) - sin(\delta_2(k,r))g(kr)\big]^2
\end{equation}
where f(kr) is 
\begin{equation}
f(kr) = \left(\frac{3}{(kr)^2}  - 1\right)sin(kr)-\frac{3}{kr}cos(kr)
\end{equation}
and g(kr) is 
\begin{equation}
g(kr) = \left(\frac{-3}{(kr)^2} + 1\right)cos(kr)- \frac{3}{kr}sin(kr)
\end{equation}
\end{enumerate}
It is important to note that solving these equations for obtaining SPS involves the potential directly, without the need for wavefunction. We have implemented the $5^{th}$ order Runge-Kutta (RK-5) method for solving these first order phase equations and optimised the model parameters by minimising the mean absolute percentage error (MAPE)-value defined as 
\begin{equation}
MAPE = \frac{1}{N} \sum_{i=1}^{N}  \frac{\lvert\delta^{exp}_i - \delta^{sim}_i\rvert} {\lvert\delta^{exp}_i\rvert} \times 100
\end{equation}
where $\delta^{exp}_i$ and $\delta^{sim}_i$ are the experimental and obtained phase-shifts. 
The implemented code is shown as a flow chart in Fig. \ref{Flowchart}.\\
%%%%%%%%%%%%%%%%%%%%%%%%%%%%%%%%%%%%%%%%%%%%%%%%%%%%%
%%%%%%%%%%%%%%%%%%%flow chart%%%%%%%%%%%%%%%%%%%%%%%%
%%%%%%%%%%%%%%%%%%%%%%%%%%%%%%%%%%%%%%%%%%%%%%%%%%%%%
Once, SPS $\delta_{\ell}(E)$ are obtained for each orbital angular momentum $\ell$, one can calculate the partial cross section $\sigma_{\ell}(E)$ \& total cross section $\sigma_{T}$ by using following formulae \cite{31} :
\begin{equation}
\sigma_{l}(E) = \frac{4\pi(2l+1)}{k^2} \sin^2({\delta_{l}(E)})
\label{Pxsec}
\end{equation} 
\begin{equation}
\sigma_{T} = \frac{4\pi}{k^2}\sum_{l}\left\lbrace (l+1)\sin^2(\delta_{l}^{+}) + l\sin^2({\delta_{l}^{-}})\right\rbrace
\label{Txsec}
\end{equation}

\section{Results \& Discussion}
The various partial waves for both $n\alpha$ ($^5He$) and $p\alpha$ ($^5Li$) systems are $s$, $p_{3/2}$, $p_{1/2}$, $d_{5/2}$ and $d_{3/2}$. The $n\alpha$ and $p\alpha$ scattering studies show sharp resonance for $p_{3/2}$ and a slightly broader resonance for first excited $p_{1/2}$ state. The experimental SPS data for all these states have been taken from Satchler et al. \cite{18}.  
\subsection{Study of nucleon-alpha system:}
The inverse potential parameters along with $MAPE$ values for $s$, $p$ and $d$-states of $n\alpha$, $p\alpha$ systems obtained using our approach are tabulated in Table \ref{Table1}. 

\begin{figure}[h!]
\caption{Simulated and experimental SPS for $s$ and $p$-states of $n\alpha$\textbf{(left)}, $p\alpha$\textbf{(right)}. The experimental data is taken from Satchler et al. \cite{18}}
\centering
\includegraphics[scale=0.27, angle = 270]{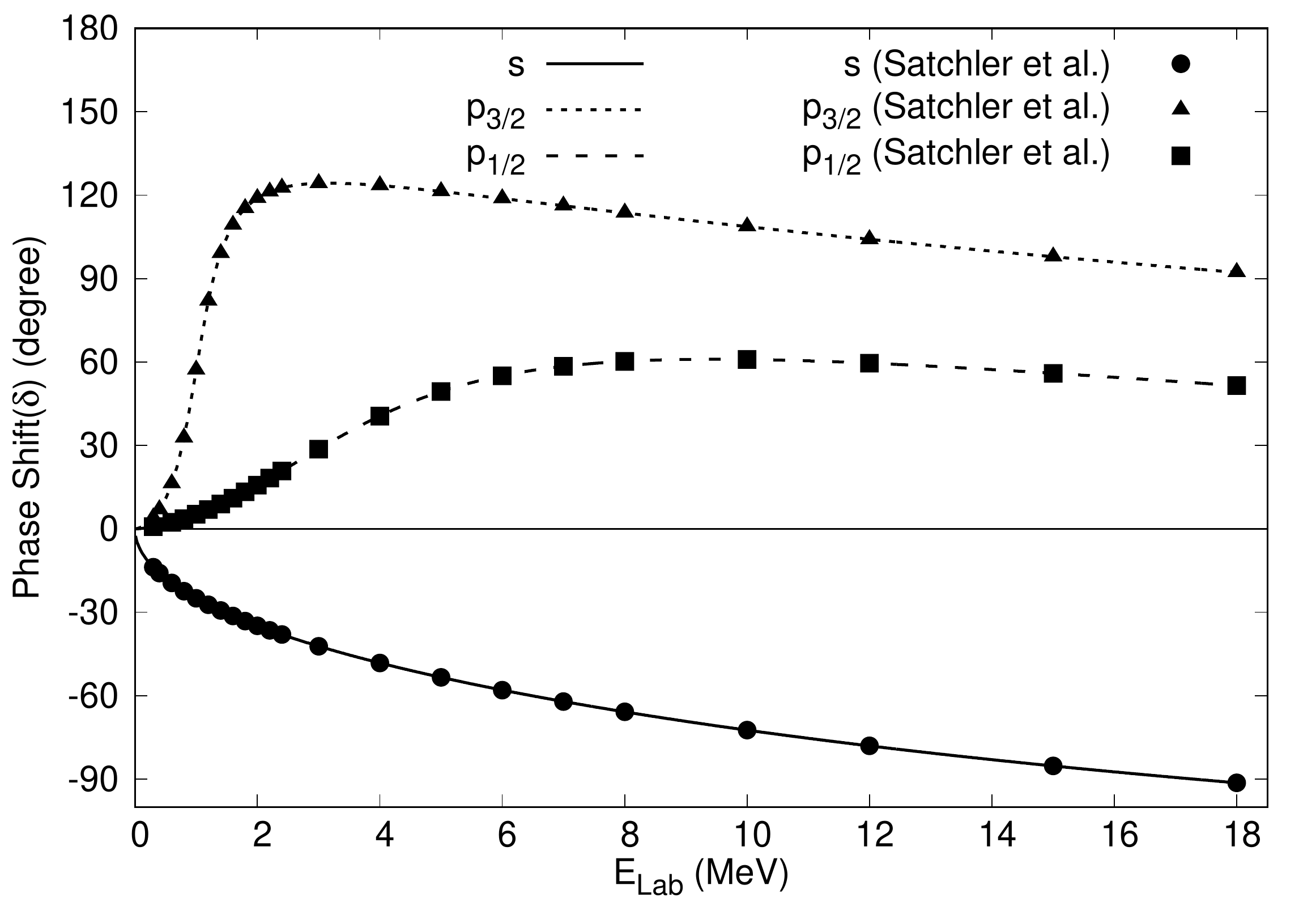}\includegraphics[scale=0.27, angle = 270]{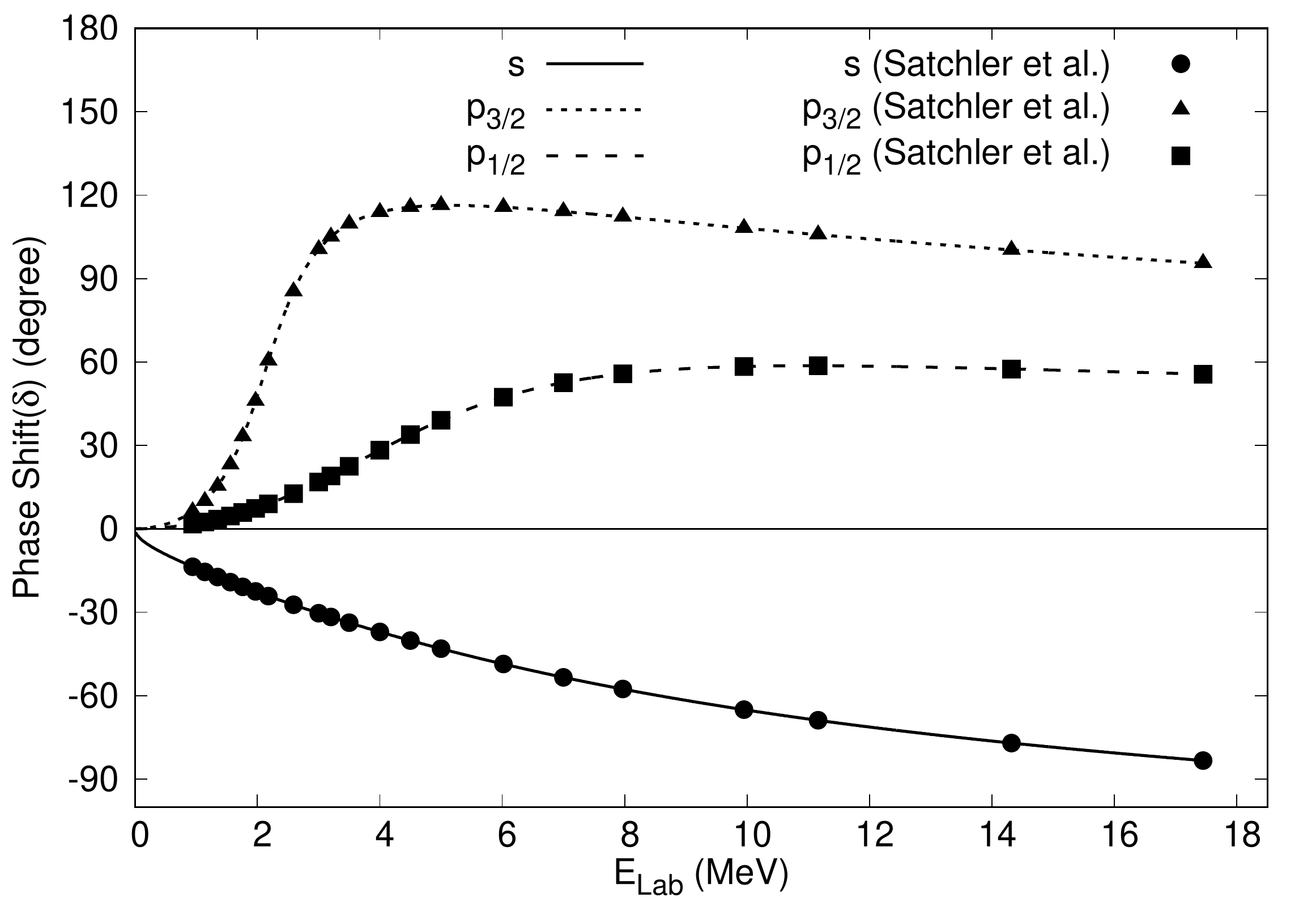}
\label{fig1} 
\end{figure}

\begin{figure}[h!]
\caption{Simulated and experimental SPS for $d$-state of $n\alpha$\textbf{(left)} and $p\alpha$\textbf{(right)}. The experimental data is taken from Satchler et al. \cite{18}}
\centering
\includegraphics[scale=0.33, angle = 270]{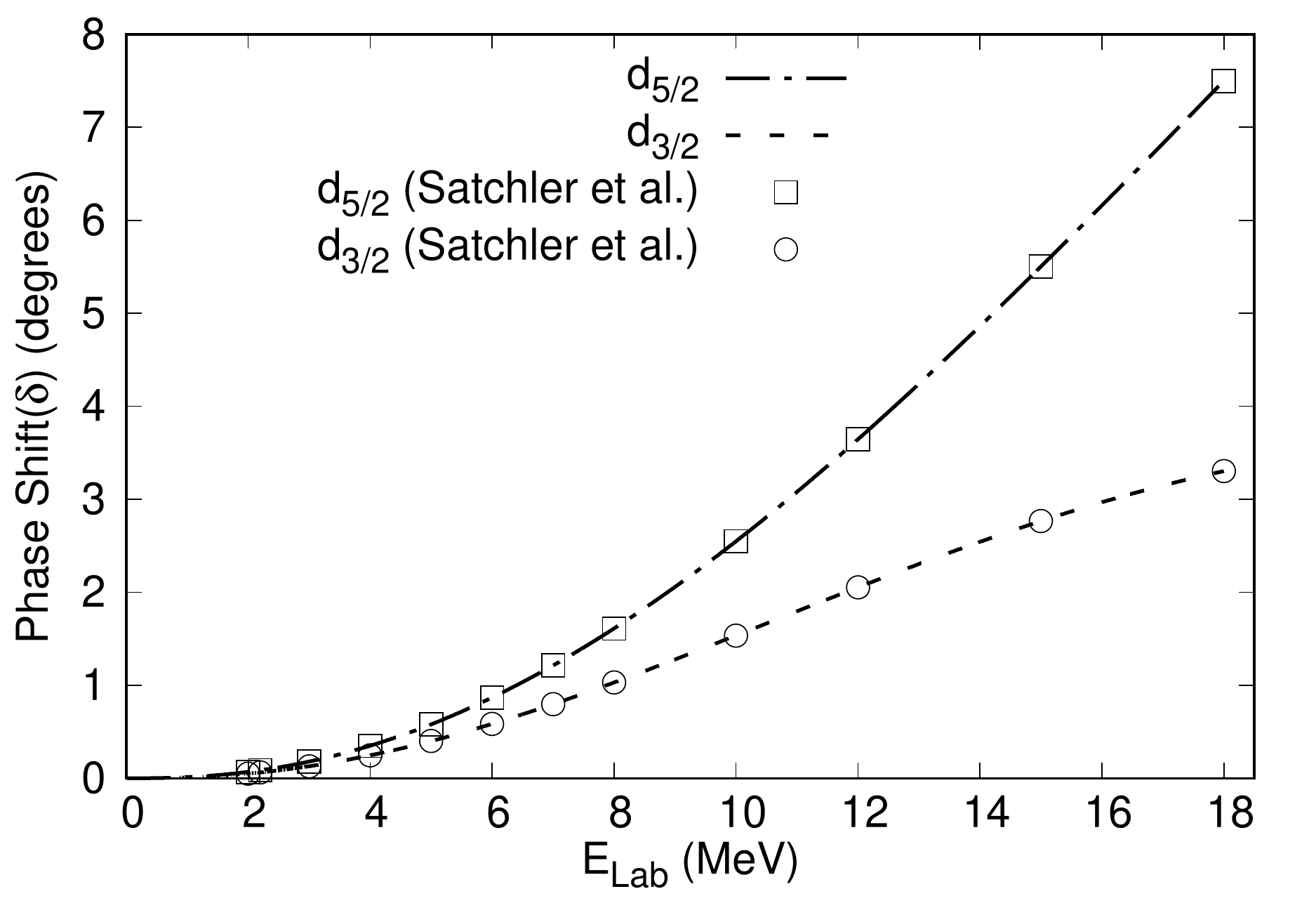}\includegraphics[scale=0.33, angle = 270]{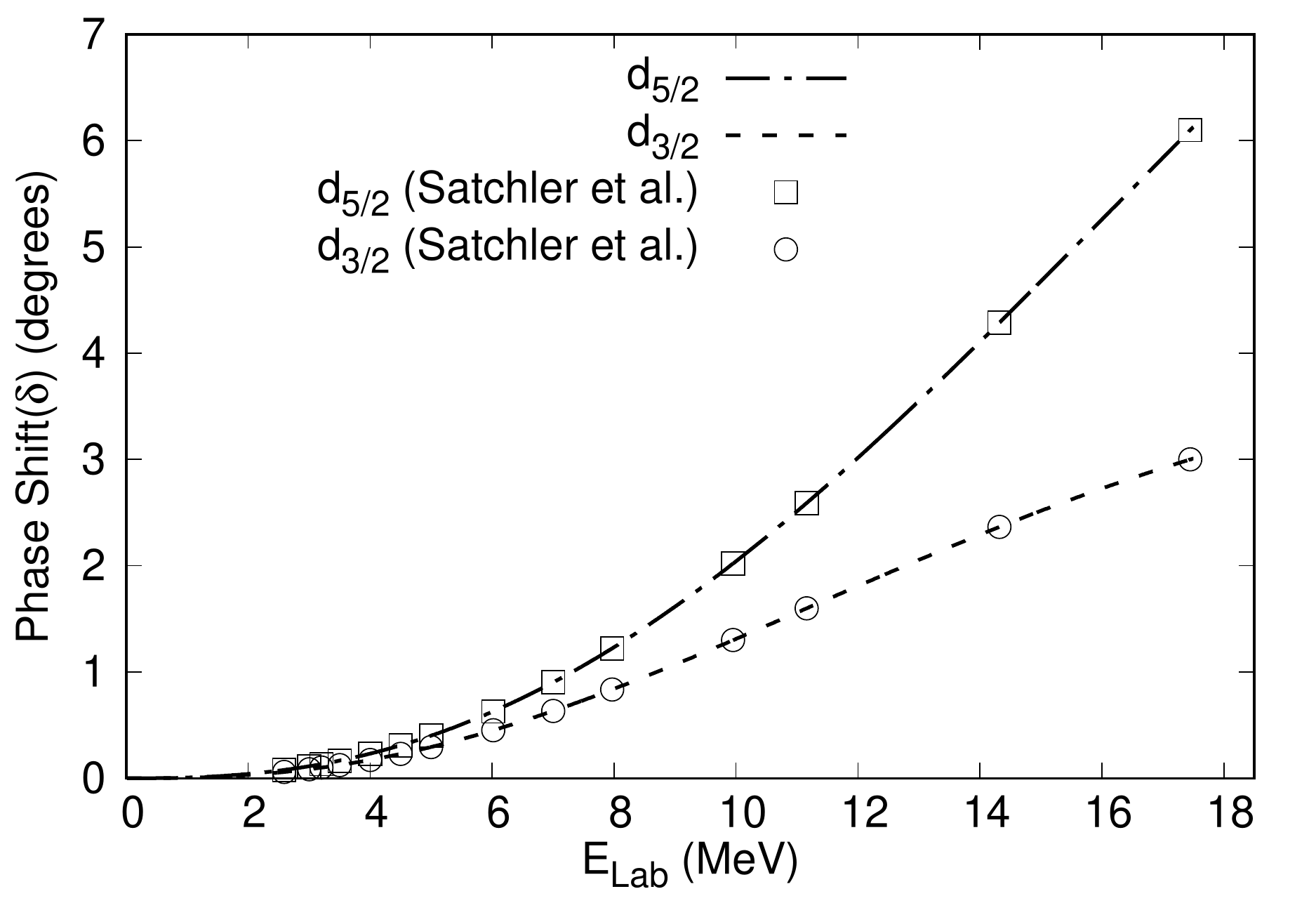} 
\label{fig2} 
\end{figure}

\begin{table}[h!]
\centering
\caption{List of model parameters of Morse potential for various systems and states along with corresponding MAPE values.}
\label{Table1}
\scalebox{0.75}{
\begin{tabular}{c|ccccc} 
\hline
\textbf{System}                     & \textbf{State}     & $\textbf{V}_0(\mathrm{\textbf{MeV}})$ & $r_m\left(\mathrm{fm}\right)$ & \textbf{$a_{m}(\mathrm{fm})$} & $\textbf{MAPE(\%)}$  \\ 
\hline
\multirow{5}{*}{\textbf{n$\alpha$}} & $\mathbf{s}$ & $37.66$                               & $1.28$                        & $0.90$                        & $0.04$               \\
                                    & $\mathbf{p}_{1 / 2}$ & $45.37$                               & $2.13$                        & $0.43$                        & $0.25$               \\
                                    & $\mathbf{p}_{3 / 2}$ & $58.30$                               & $1.63$                        & $0.50$                        & $0.14$               \\
                                    & $\mathbf{d}_{3 / 2}$ & $10.58$                               & $2.54$                        & $0.42$                        & $1.53$               \\
                                    & $\mathbf{d}_{5 / 2}$ & $20.22$                               & $1.82$                        & $0.50$                        & $1.26$               \\ 
\hline
\multirow{5}{*}{\textbf{p$\alpha$}} & $\mathbf{s}$ & $33.24$                               & $1.69$                        & $1.22$                        & $1.18$               \\
                                    & $\mathbf{p}_{1 / 2}$ & $22.83$                               & $1.61$                        & $0.84$                        & $1.95$               \\
                                    & $\mathbf{p}_{3 / 2}$ & $65.22$                               & $1.02$                        & $0.66$                        & $1.23$               \\
                                    & $\mathbf{d}_{3 / 2}$ & $3.95$                                & $0.44$                        & $2.04$                        & $2.17$               \\
                                    & $\mathbf{d}_{5 / 2}$ & $28.25$                               & $0.21$                        & $1.07$                        & $1.45$               \\ 
\hline
{\textbf{nd}}        & $^{2}S_{1/2}$            & $173.29$ &	$2.53$ &	$0.89$                            & $1.95$               \\ % 
\hline
{\textbf{pd}}        & $^{2}S_{1/2}$             & $33.46$	&	$2.09$ &	$1.33$ & $2.05$               \\
\hline
\end{tabular}}
\end{table}
%%%%%%%%%%%%%%%%%%%%%%%%%%%%%%%%%%%%%%%%%%%%%%%%%%%%%%%%%%%%%%%%%%%%%%%%%%%%%%%%%%%%%%%
The obtained SPS for $s$, $p_{1/2}$ and $p_{3/2}$-states along with experimental data \cite{18} are shown in Fig. \ref{fig1} and a good agreement can be clearly seen. It should be noted that for sake of convienience, SPS plot for $s$-state is shifted by $180^{o}$ to avoid overlapping with other SPS plots.
%%%%%%%%%%%%%%%%%%%%%%%%%%%%%%%%%%%%%%%%%%%%%%%%%%%%%%%%%%%%%%%%%%%%%%%%%%%%%%%%%%%%%
The SPS plots for $d_{3/2}$ and $d_{5/2}$ are plotted separately in Fig. \ref{fig2} to bring out their match with experimental data clearly, because the SPS for these are small as compared to other states. Once again, the match between obtained \& experimental SPS is very good.  
%%%%%%%%%%%%%%%%%%%%%%%%%%%%%%%%%%%%%%%%%%%%%%%%%%%%%%%%%%%%%%%%%%%%%%%%%%%%%%%%%%%%%%%%%%%%%%%%%
\begin{figure}[!h]
\caption{Interaction Potentials for $s$, $p$ and $d$-states of $n\alpha$\textbf{(left)} and $p\alpha$\textbf{(right)} along with centrifugal term.}
\centering
\includegraphics[scale=0.33,angle = 270]{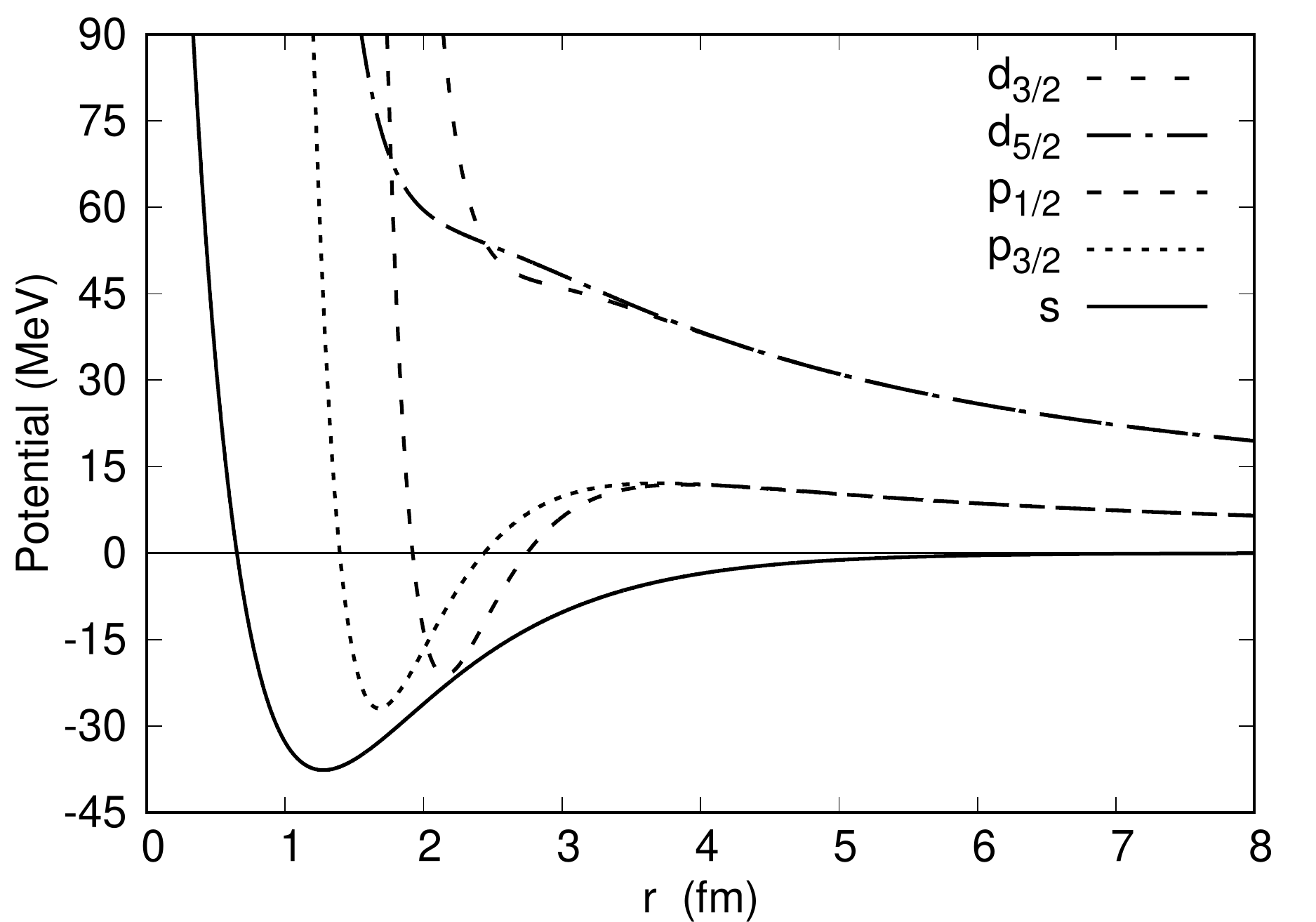}\includegraphics[scale=0.33,angle = 270]{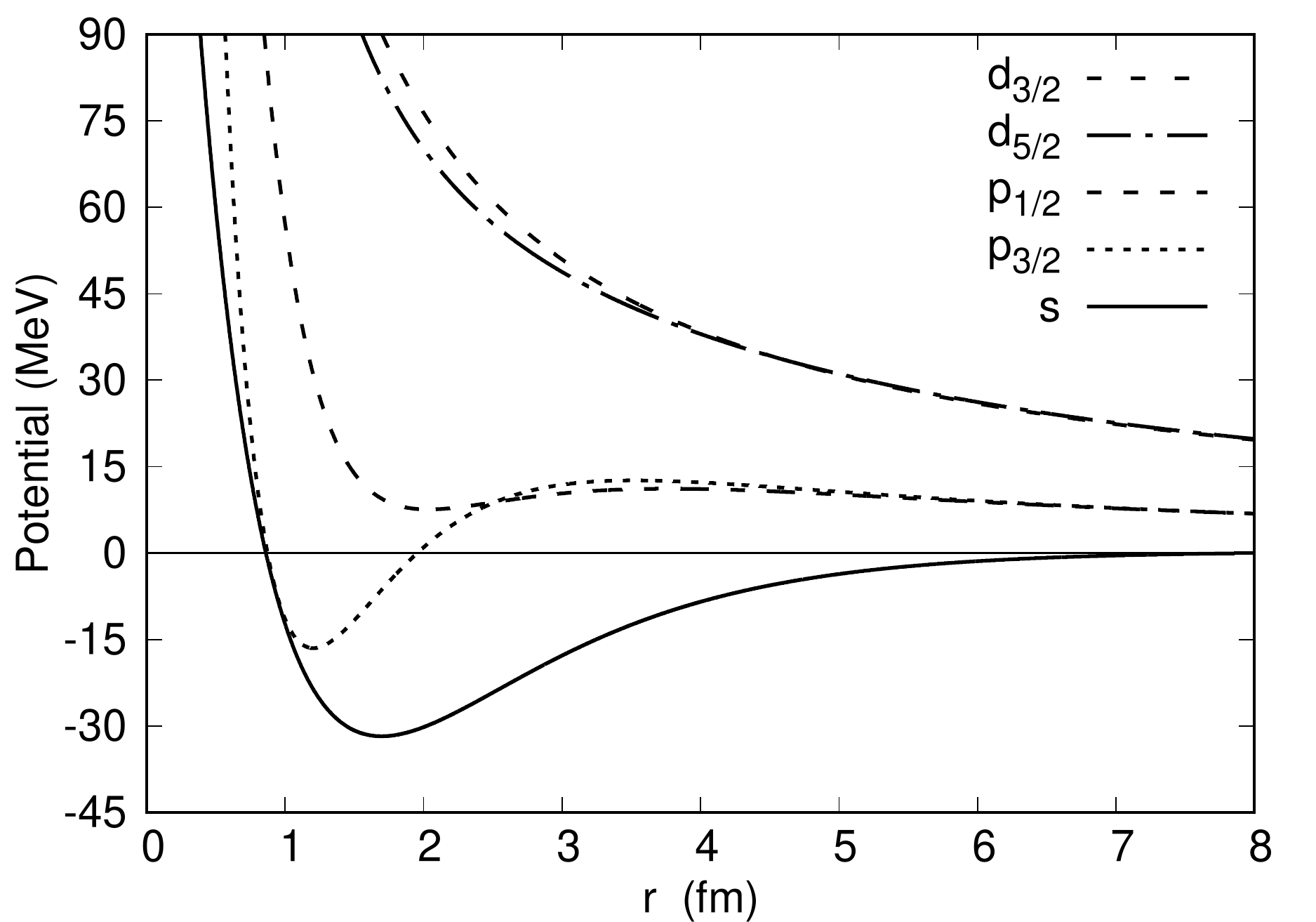} 
\label{fig33} 
\end{figure}
%%%%%%%%%%%%%%%%%%%%%%%%%%%%%%%%%%%%%%%%%%%%%%%%%%%%%%%%%%%%%%%%%%%%%%%%%%%%%%%%%%%%%%%%%%%%%%%%%
In Fig. \ref{fig33}, the total interaction potential  including that of centrifugal term are plotted for all the five states for both systems. It is clearly seen that, the potentials of $n\alpha$ are slightly deeper in comparison to corresponding states in $p\alpha$ system owing to the Coulomb term in the later.
Using obtained SPS, both partial cross section $\sigma_{l}(E)$ and total cross section $\sigma_{T}$ for $n\alpha$, $p\alpha$ elastic scattering are calculated for $E_{lab}$ values up to 18 MeV using Eqs.\ref{Pxsec}-\ref{Txsec}. The partial cross section for $p$ states are shown in Fig.\ref{fig5}. 
For $n\alpha$ the resonance peaks in center of mass energy for $p_{1/2}$ and $p_{3/2}$ are observed(experimental) at $4.1 (4 \pm 1)$ MeV and $0.93 (0.89)$ MeV and their respective decay widths are $8.19 (4 \pm 1)$ MeV and $0.88 (0.60)$ MeV. And for p-$\alpha$ the resonance peaks in center of mass energy for $p_{1/2}$ and $p_{3/2}$ are observed(experimental) at $5.21 (5-10)$ MeV and $1.96 (1.96)$ MeV and their respective decay widths are $9.54 (5 \pm 2)$ MeV and $1.90 (1.5)$ MeV \cite{38,39,40}. 
\begin{figure}[!ht]
\caption{Plots of partial cross section w.r.t $E_{Lab}$ energies of resonant states $p_{1/2}$ and $p_{3/2}$. for $n\alpha$\textbf{(top)} and $p\alpha$\textbf{(bottom)} systems.}
\centering
\includegraphics[scale=0.43,angle = 360]{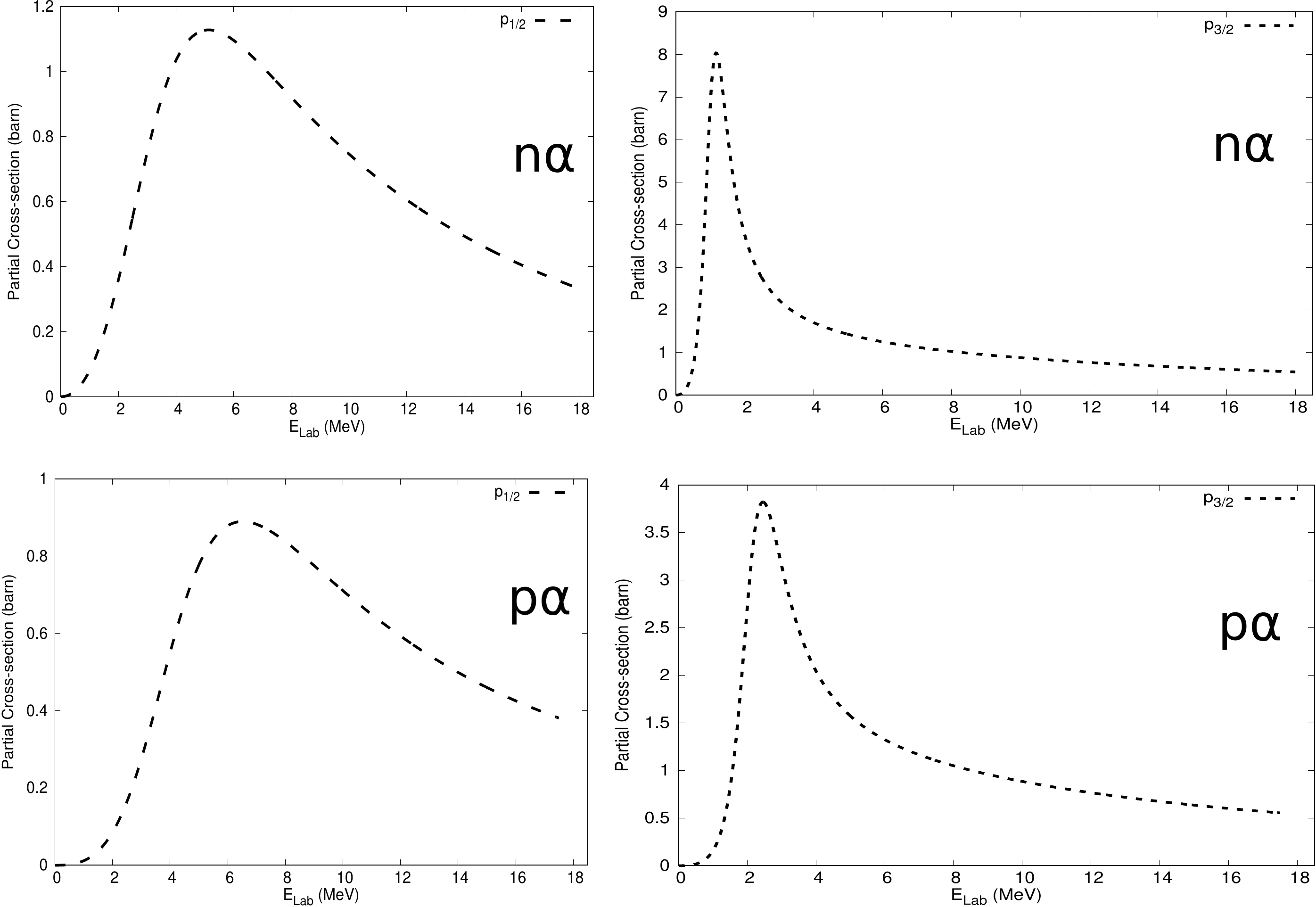}
\label{fig5} 
\end{figure}

\begin{figure}[h!]
\caption{Total cross-section ($\sigma_{T}$) for $n\alpha$\textbf{(left)} and $p\alpha$\textbf{(right)} systems. Continuous line is from our work and empty black circles are from \cite{41}.}
\centering
\includegraphics[scale=0.33,angle = 270]{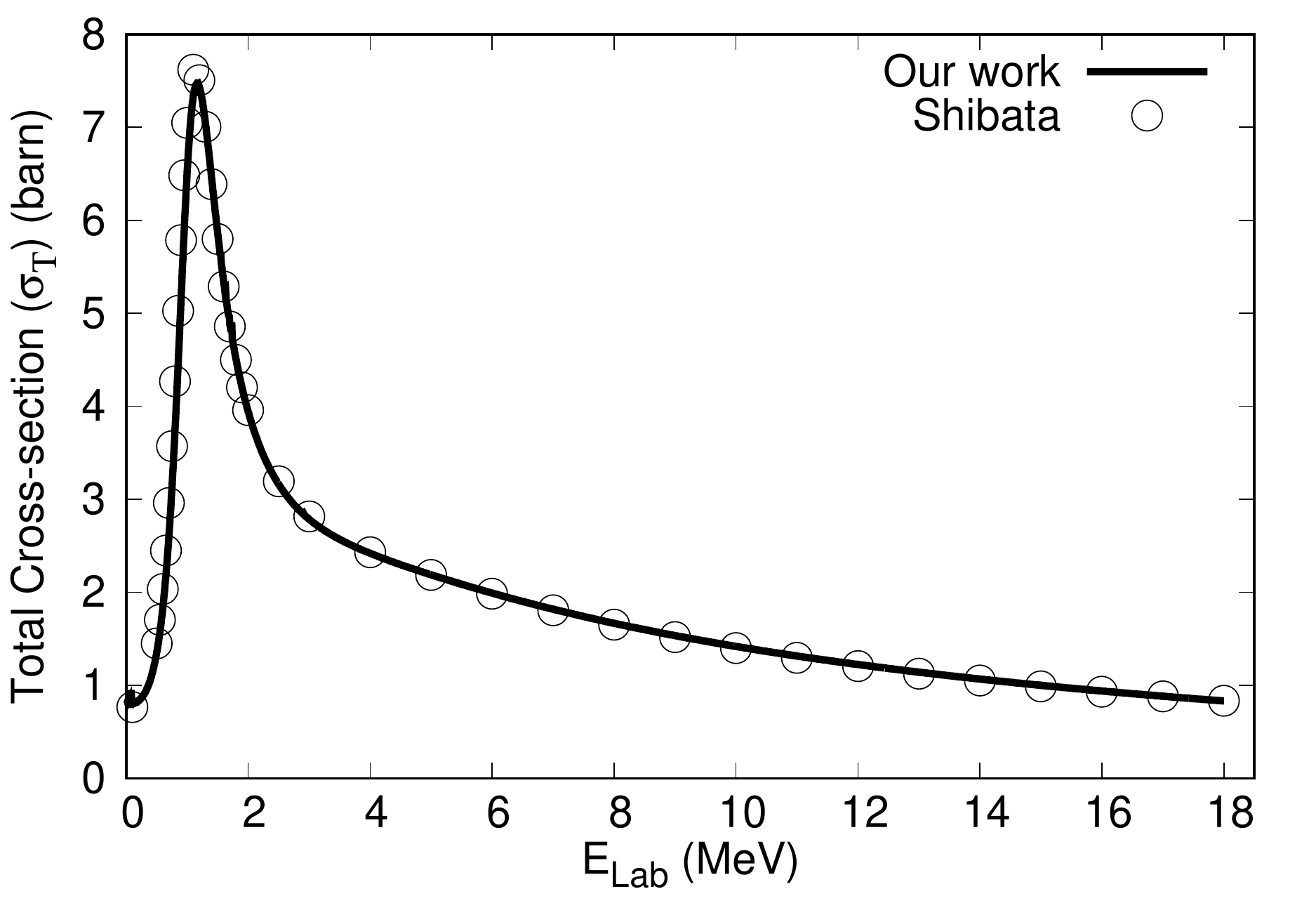}\includegraphics[scale=0.33,angle = 270]{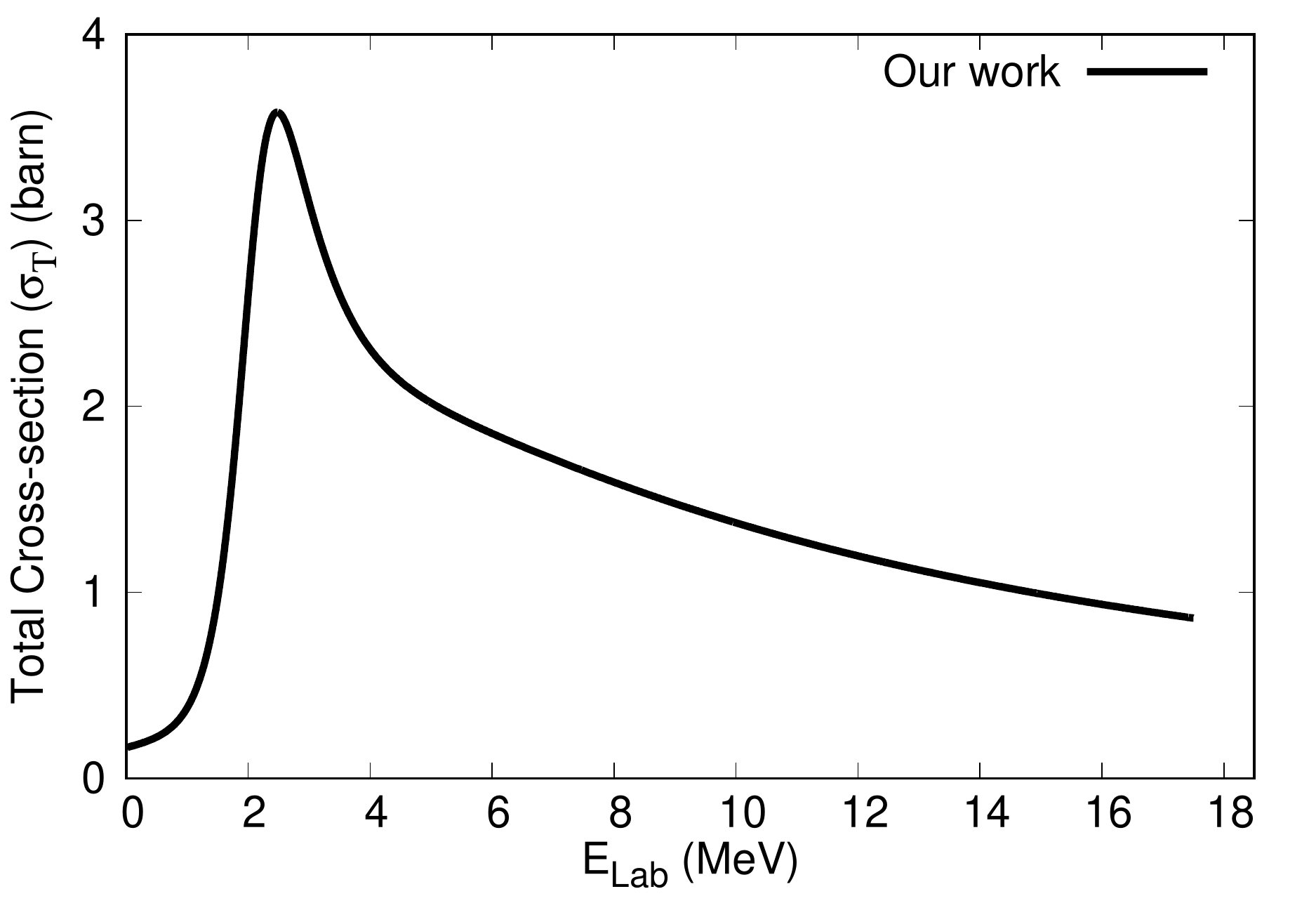}  
\label{fig6} 
\end{figure}
%%%%%%%%%%%%%%%%%%%%%%%%%%%%%%%%%%%%%%%%%%%%%%%%%%%%%%%%%%%%%%%%%%%%%%%%%%%%%%%%%%%%%%%%%%%
\subsection{Study of nucleon-deuteron systems:}
\begin{figure}[h!]
\caption{SPS\textbf{(left)} and interaction potential\textbf{(right)} for $nd$ system.}
\centering
\includegraphics[scale=0.45,angle = 270]{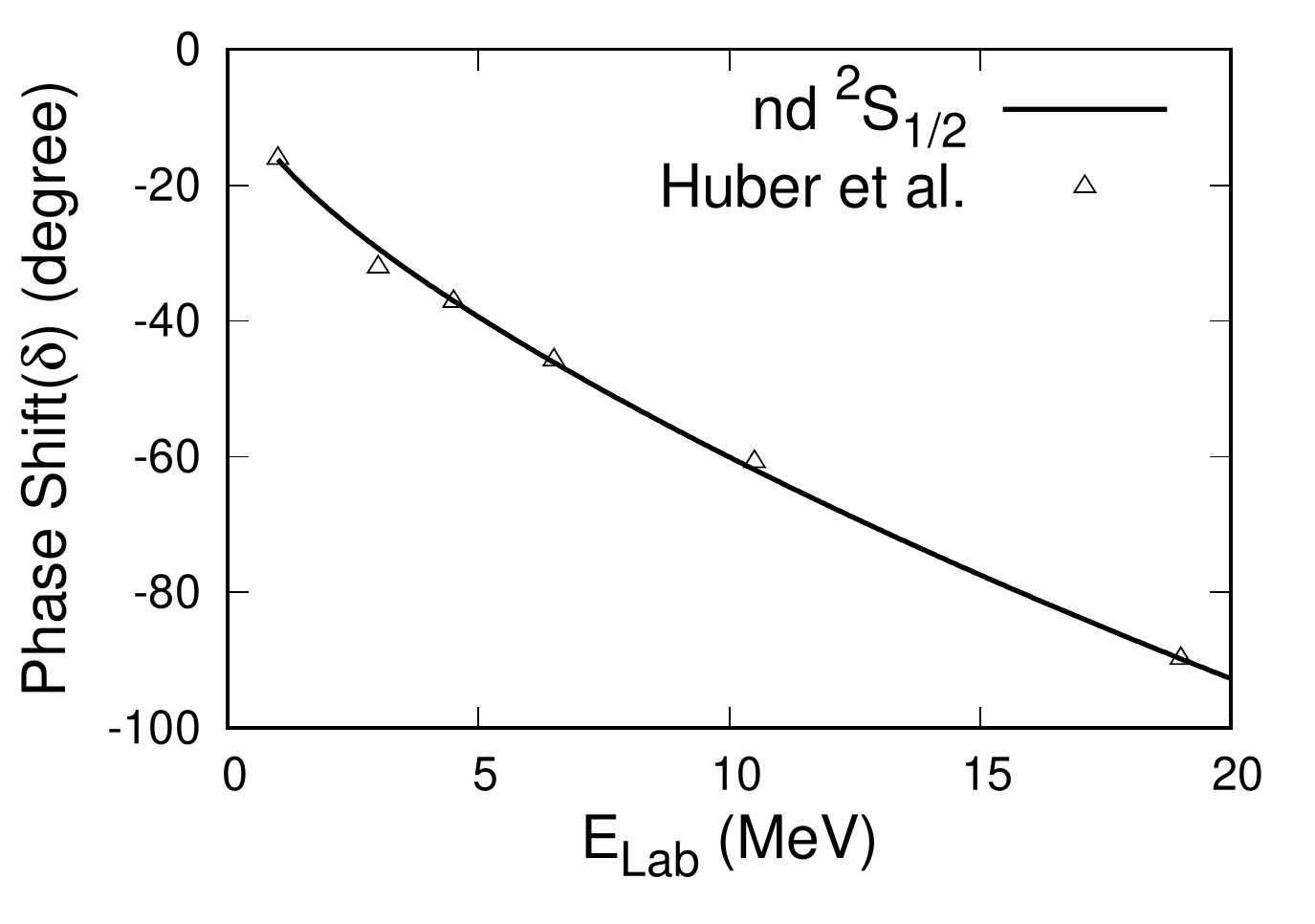}\includegraphics[scale=0.45,angle = 270]{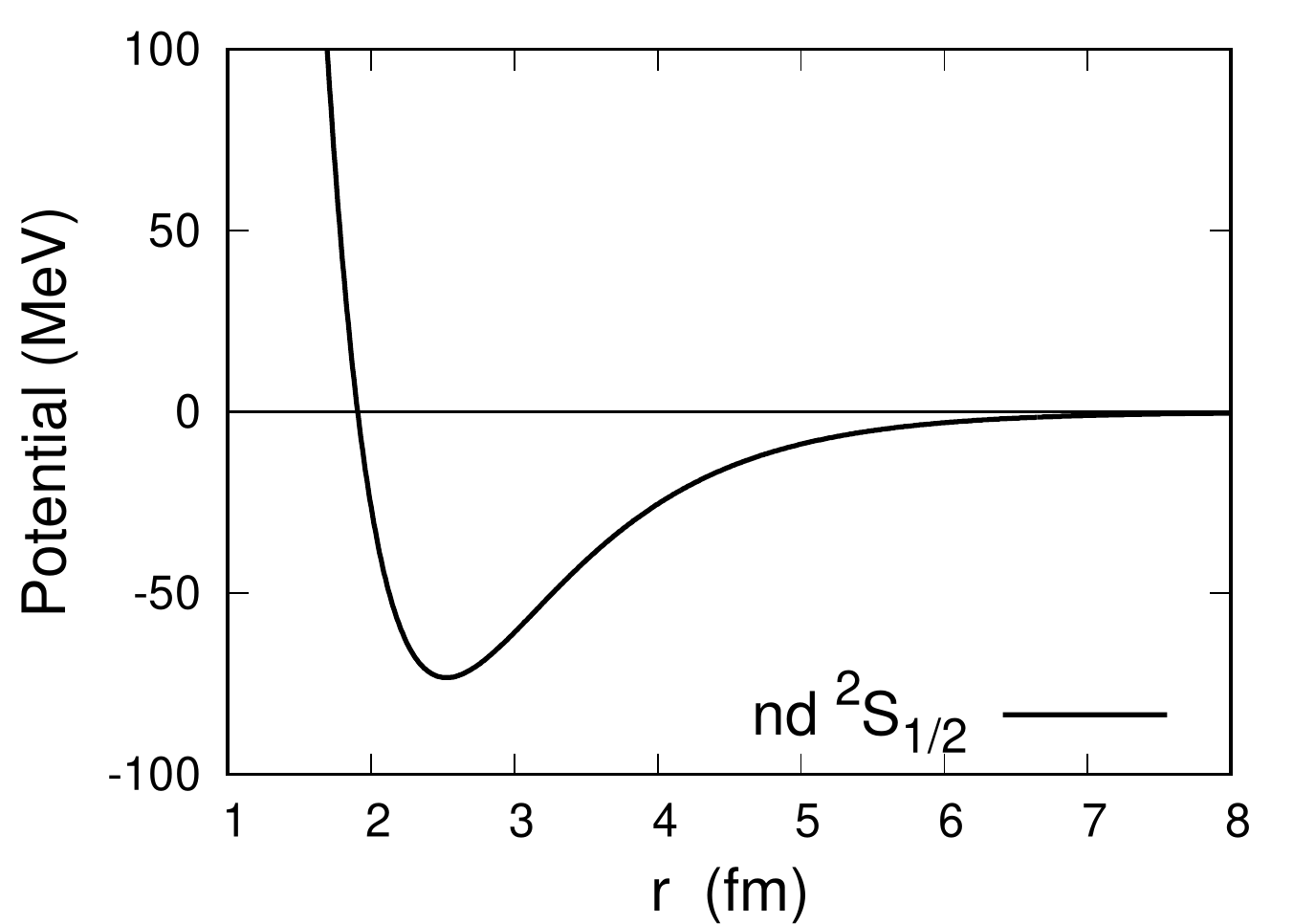}
\label{fig7} 
\end{figure}

\begin{figure}[h!]
\caption{SPS\textbf{(left)} and interaction potential\textbf{(right)} for $pd$ system.}
\centering
\includegraphics[scale=0.45,angle = 270]{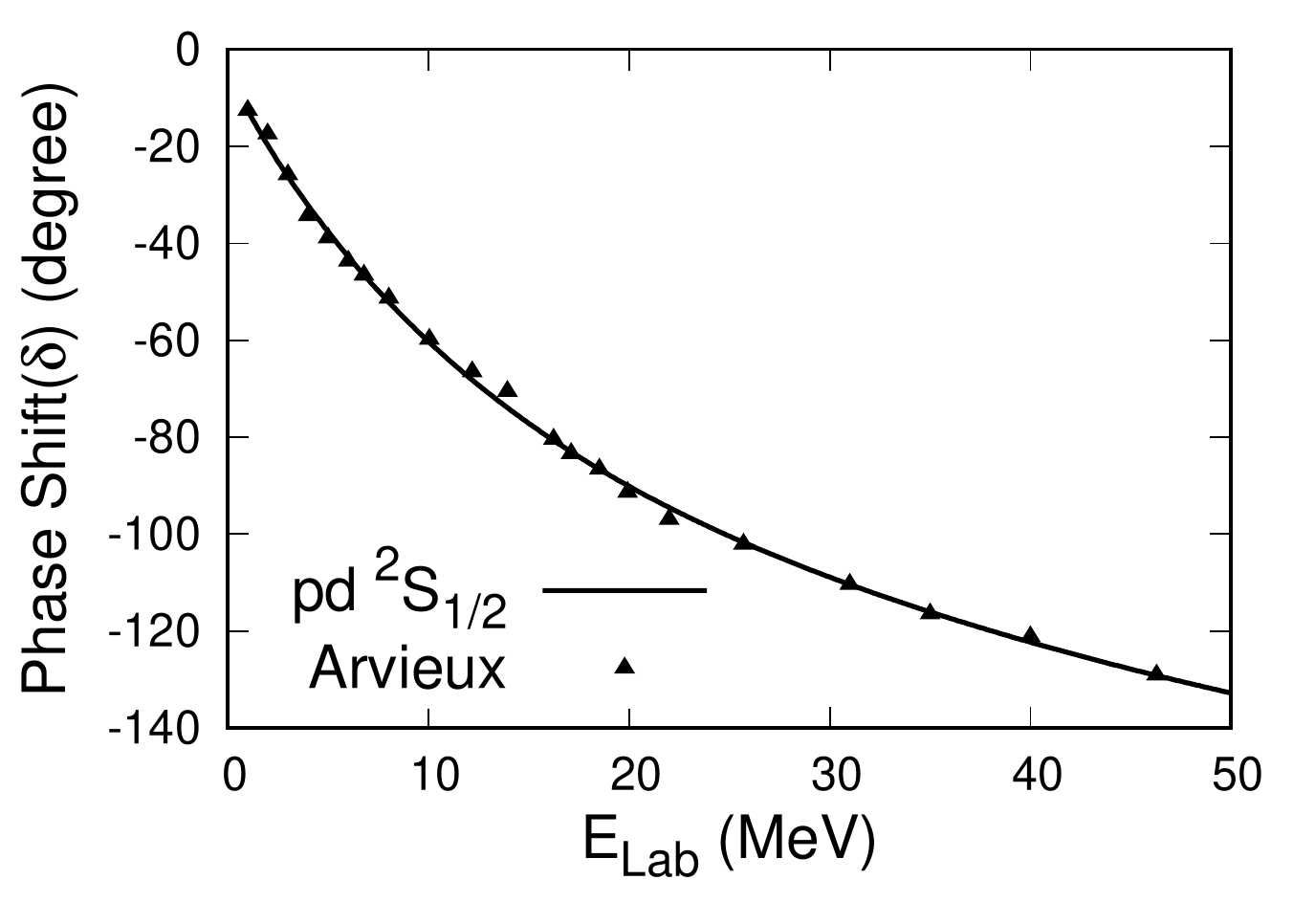}\includegraphics[scale=0.45,angle = 270]{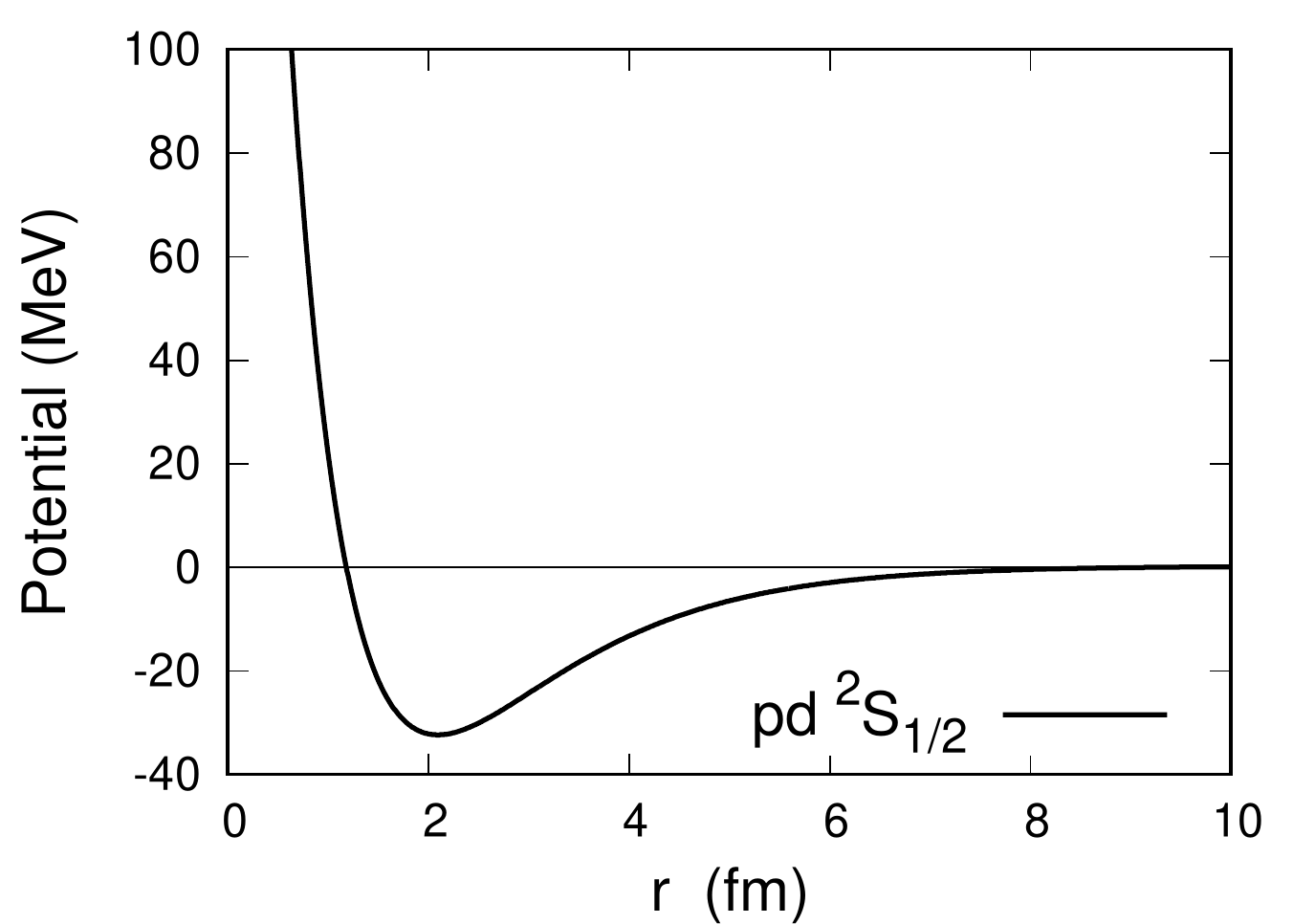}  
\label{fig8} 
\end{figure}

An extensive analysis on elastic nucleon-deuteron scattering has been done by Huber $et al.$ \cite{1} and J. Arvieux \cite{2} separately for laboratory energies ranging from $1-19$ $MeV$ for \textit{nd} and $1-46.3$ $MeV$ for \textit{pd} and these are considered as standard data for these systems. The model parameters for Morse potential obtained by using PFM for $^2S_{1/2}$  ground state of \textit{nd} and \textit{pd }scattering have been tabulated in Table \ref{Table1}. The mean absolute percentage error (MAPE) has also been given along with.\\
In Fig. \ref{fig7}, the obtained SPS are compared with experimental data \cite{1} on the left side and the corresponding inverse potential for the doublet S-wave ground state of \textit{nd} is shown.  The potential is attractive in nature even though the experimental phase shifts are negative. This is because, for \textit{nd} system an extra phase shift of $180^{o}$ is added \cite{3} while determining the inverse potential. Similarly, the SPS and inverse potential for \textit{pd} system are shown Fig. \ref{fig8} (left) and Fig. \ref{fig8} (right) respectively. Once again, the depth of \textit{pd} system is observed to be less than that of \textit{nd} owing to the extra contribution due to Coulomb potential.

\section{Conclusion}
%The n-$\alpha$ and p-$\alpha$ elastic scattering has been modeled using Morse function as an interaction potential and the scattering phase shifts have been obtained using the phase function method. The computed phase shifts are found to be in good agreement with that of Satchler et al. In addition, resonance energy $E_{r}$ and decay width $\Gamma$ for resonant states $p_{3/2}$ and $p_{1/2}$ have been obtained from their partial cross-section determined from the computed SPS. The final result is shown in Table \ref{Table44}.
The inverse potentials for \textit{nd}, \textit{pd}, $n\alpha$, $p\alpha$ systems have been obtained using  Morse function as reference. The model parameters are optimised to reduce the mean absolute percentage error between phase shift obtained, from solving phase equation using RK-5 method, and experimental data. The computed phase shifts for \textit{nd} and \textit{pd} systems have been found to be closely matching with those of \textit{Huber} \& \textit{Arvieux} respectively. In case of $n\alpha$ and $p\alpha$ systems, not only are the scattering phase shifts matching well with those of \textit{Satchler}, but the resonance frequencies for the first two scattering $p$-states for both systems are in excellent match with the experimental data as compared in Table \ref{Table44}.   
\begin{table}[h!]
\centering
\caption{Resonance parameters for $^{5}He$ and $^{5}Li$ determined from partial cross section plots.}
\label{Table44}
\scalebox{0.7}{
\begin{tabular}{ccccccccc} 
\toprule
\multirow{2}{*}{\begin{tabular}[c]{@{}c@{}}\textbf{State}\\\end{tabular}} & \multicolumn{2}{c}{$^{5}He$ \cite{42}} & \multicolumn{2}{c}{$^{5}He$ (Our)}     & \multicolumn{2}{c}{$^{5}Li$ \cite{42}}                          & \multicolumn{2}{c}{$^{5}Li$ (Our)}                                              \\ 
\cline{2-9}
                                                                          & $E_r (MeV)$ & $\Gamma_{c.m.} (MeV)$               & $E_r (MeV)$ & $\Gamma_{c.m.} (MeV)$ & \multicolumn{1}{c}{$E_r (MeV)$} & \multicolumn{1}{c}{$\Gamma_{c.m.} (MeV)$} & \multicolumn{1}{c}{$E_r (MeV)$} & \multicolumn{1}{c}{$\Gamma_{c.m.} (MeV)$}  \\ 
\hline
$p_{1/2}$                                                             & $4\pm 1$    & $4\pm 1$                        & 4.1         & 8.19                & 5-10                            & $5\pm 2$               & 5.21                            & 9.54                                     \\
$p_{3/2}$                                                             & 0.89        & $0.60\pm 0.02$                    & 0.93        & 0.88                & 1.96                            & $\approx 1.5$          & 1.96                            & 1.90                                     \\
\bottomrule
\end{tabular}}
\end{table}

\clearpage
%\section*{References}

\end{document}